\documentclass{sig-alternate-05-2015}

\usepackage{tikz}
\usetikzlibrary{calc}
\usepackage{url}
\usepackage{mdframed}
\usepackage{algpseudocode}
\usepackage{pgfplots}
\usepackage{pgfplotstable}
\usepackage{graphicx}
\usepackage{subcaption}
\usepackage{caption}
\usetikzlibrary{calc}
\usetikzlibrary{patterns}
\usetikzlibrary{shapes,snakes}
\usetikzlibrary{shapes.multipart, arrows}
\usetikzlibrary{decorations.pathreplacing}
\usepackage{multirow}
\usepackage{balance}
\usepackage{algorithm,algpseudocode}
\usepackage{amsmath} 
\usepackage{xcolor,colortbl}
\usepackage{enumitem}
\usepackage{tabularx}

\usepackage{pdflscape}
\usepackage{multicol}

\makeatletter
\def\@copyrightspace{\relax}
\makeatother

\def\codename{\textproc{AMES}}
\def\heartbeatexchange{\textproc{HeartbeatExchange}}
\def\newinstance{\textproc{NewProcess}}
\def\failedinstance{\textproc{ExpelProcess}}
\def\shutdown{\textproc{ShutDown}}

\begin{document} \sloppy

\definecolor{cornellred}{rgb}{0.7, 0.11, 0.11}
\definecolor{arsenic}{rgb}{0.23, 0.27, 0.29}
\definecolor{auburn}{rgb}{0.43, 0.21, 0.1}
\definecolor{charcoal}{rgb}{0.21, 0.27, 0.31}
\definecolor{deepcarrotorange}{rgb}{0.91, 0.41, 0.17}
\definecolor{eggplant}{rgb}{0.38, 0.25, 0.32}

\title{Autonomous Membership Service for Enclave Applications}

\author{{Hung Dang, Ee-Chien Chang}\\
School of Computing, National University of Singapore\\
\{hungdang,changec\}@comp.nus.edu.sg\\
}

\maketitle

\section*{Abstract}

Trusted Execution Environment, or enclave, promises to protect
data confidentiality and execution integrity of an outsourced computation on an untrusted host. Extending the protection to distributed applications that run on physically separated hosts, however, remains non-trivial. For instance, the current enclave provisioning model hinders elasticity of cloud applications. 
Furthermore, it remains unclear how an enclave process could verify if there exists another
concurrently running enclave process instantiated using the same codebase, or count a number of such processes.
In this paper, we seek an autonomous membership service for enclave applications. The application owner only needs to partake in instantiating the very first process of the application, whereas all subsequent process commission and decommission will be administered by existing and active processes of that very application.
To achieve both safety and liveness, our protocol design admits unjust excommunication of a non-faulty process from the membership group. 
We implement the proposed membership service in a system called \codename. Our experimental study shows that \codename\ incurs an overhead of $5\% - 16\%$ compared to vanilla enclave execution. 
\section{Introduction}
\label{sec:intro}
Intel Software Guard Extensions (SGX)~\cite{sgx} protects data confidentiality and execution integrity of outsourced computations by offering a hardware-protected trusted execution environment (TEE), or \textit{enclave}, that guards the user's security-critical application code and its data against untrusted system software including the operating system (OS). The CPU prevents any non-enclave code from reading or modifying enclave memory at runtime. SGX architecture relies on the OS to instantiate enclave. A user can verify if a specific enclave is correctly instantiated and running at a remote host via a \textit{remote attestation} protocol~\cite{sgx_remote_attest}, which also establishes a secure channel via which the user can provision application secrets, if any, to the enclave.

While SGX allows users to harden security of software applications on an untrusted host (or server)~\cite{haven, pdedup}, extending its protection to distributed applications running on a number of hosts is non-trivial. More specifically, if the application involves a secret that must not be revealed to the untrusted host, the secret provisioning requires remote attestation by the application owner. 
In the context of cloud service, this hinders the elasticity of cloud applications, for the cloud service providers cannot commission additional enclave processes dynamically at its sole discretion.
In another example, it is unclear how an enclave process could verify if there exists another concurrently running enclave process instantiated using the same codebase~\cite{sgx_singleton_matter}, or count the number of such processes, in an autonomic manner. This capability is desired in implementing singleton applications (i.e., only a single instance of the application can be created and run) and decentralized floating licensing (i.e., floating licensing without the central license server).

Let us use an example of a \textit{forking attack}~\cite{forking_attacks} on application that maintains persistent state to illustrate the challenges of extending TEE protections to a distributed environment. At a high level, the adversary (e.g., cloud service provider) instantiates multiple enclave processes of the same application and feeds them with different, potentially conflicting, inputs so as to violate the expected behaviours of the victim application. 
More specifically, consider an application $\texttt{Prog}$ that serves read and write requests of a stateless client. The adversary runs two independent enclave processes (or processes for short) of $\texttt{Prog}$, denoted by $p_1$ and $p_2$, while keeping them oblivious to each other's
existence. Since the adversary handles all I/O operations to the processes, the client is only presented with a simple API for read and write requests, and is unaware of the underlying provisioning of the processes. At the beginning, both processes assume an initial state $s_0$. The client issues two sequential requests, the first writes data $\texttt{dat}$ to $\texttt{Prog}$'s memory, changing its state to $s_1$, and the second reads from $\texttt{Prog}$ its latest written data. From the application logic, the client should receive $\texttt{dat}$ as a response to the second request. Nevertheless, if the adversary routes the write request to $p_1$ and read request to $p_2$, the client will receive an obsolete (but authenticated) response drawn from the stale state $s_0$. This attack may wreak severe havoc in financial applications wherein an account balance should reflect at all time a correct transaction history. 

The example described earlier motivates the need of a \textit{membership service} for enclave applications in a distributed environment. The membership service of an application $\texttt{Prog}$ not only facilitates seamless commissioning of $\texttt{Prog}$'s enclave processes, but also monitors and keeps track of the status of all $\texttt{Prog}$'s active processes, ensuring that their collective existence adheres to the Service Level Agreement (SLA). In a straight forward approach, one can implement the membership service by maintaining a special \textit{directory server} that keeps track of all active processes, and excommunicates a process that it suspects of being faulty from the membership group. The commissioning of a new process is augmented such that the newly instantiated process remains \textit{nonoperational} until its request to join the existing membership group is approved by the directory server. Nonetheless, this approach is not desirable, for the directory server becomes an obvious single-point-of-failure. A logical attempt to avoid the single-point-of-failure would be to replicate the directory server into a distributed fault-tolerant directory service comprising of a cluster of replicas coordinating their operations via a consensus protocol such as Paxos~\cite{paxos}. However, it is challenging to protect these replicas themselves against forking attack~\cite{forking_attacks}, wherein the adversary manage to splits the directory service replicas into two or more ``cliques'', each of which is oblivious to the others' existence. Such adversary can then selectively route communications of the existing processes, or join request of the newly instantiated processes to different clique of directory service replicas, thereby evading the control mechanism that enforces the SLA.

In this paper, we seek a membership service that is autonomous. In particular, once an application  $\texttt{Prog}$  has been successfully bootstrapped, all subsequent process commission and decommission  of $\texttt{Prog}$ are administered by its own active processes.
This self-administering is attained by designating one of $\texttt{Prog}$'s active processes at a {\it leader}, while the other assume the role of follower. The leader is tasked to monitor the status of all follower processes by exchanging heartbeat messages. 
On the one hand, the process commission requires acknowledgement from the leader and a quorum of existing followers before the new process becomes fully functional. On the other hand, any process that is suspected of having halted is excommunicated from the membership group. 
Our protocol design necessarily admits unjust suspicion and excommunication of a non-faulty process, in order to sidestep the difficulty of solving the membership problem in an asynchronous environment when faults may be present~\cite{transis}. To prevent the non-faulty yet excommunicated process from disrupting the membership service, its future communication to other processes of $\texttt{Prog}$ is discarded, rendering it isolated. The isolated process eventually halts by committing suicide, fulfilling the excommunication.

We consider a powerful adversary that controls the OS (or a hypervisor) of the hosts on which the enclave processes are instantiated. The adversary can schedule or corrupt all non-enclave processes, and restart the processes at will. Further, it can control all network communications between the processes regardless of whether they are instantiated on the same or across different hosts. The adversary can modify, reorder or delay the network messages arbitrarily. Nonetheless, we assume that the adversary cannot break enclave protections, and is limited by the constraints of the cryptographic methods employed.


We implement the proposed solution that enables autonomous membership service for SGX enclaves in a system called \codename. The application owner only needs to partake in the commissioning of the first enclave process (i.e., conducting remote attestation and provision application secret to the enclave). All subsequent process commissioning (or dismantling) during the lifetime of the application will be handled automatically and securely. 
To facilitate the incorporation of \codename\ logic to an enclave application $\texttt{Prog}$,
we provide application developers with a \codename\
library that envelops the membership service's operations
in a control thread, separating them from the original operational logic of $\texttt{Prog}$
that is to be implemented in a worker thread (or threads).

In summary, we make the following contributions in this paper.
\begin{itemize}
\itemsep2mm 
\item We propose a new approach to enable autonomous membership service for enclave-based applications, which allows seamless commissioning and dismantling of processes over the lifetime of the application. 

\item We provide a \codename\ library that encapsulates the membership service's operations, which can be easily incorporated to any enclave-based application. 

\item We conduct empirical evaluation of the overhead incurred by \codename\ on both our local cluster with 35 servers and the Google Cloud Platform (GPC) nodes spanning across 4 regions. The experimental results show that \codename\ incurs an overhead of $5\%-16\%$ over vanilla enclave execution.

\end{itemize}

The rest of this paper is organised as follows. 
Section~\ref{sec:background} provides backgrounds on key properties of Intel SGX and distributed consensus protocols. Section~\ref{sec:problem_statement} presents our problem statement, covering system and adversarial models, design goals, and challenges. We propose our autonomous membership service and its rationale in Section~\ref{sec:design}, before discussing the implementation details in  Section~\ref{sec:implementation}. Section~\ref{sec:eval} reports our empirical evaluation, while Section~\ref{sec:related_work} reviews the related work. Finally, Section~\ref{sec:conclusion} concludes our work.

\section{Preliminaries}
\label{sec:background}
In this section, we first describe key characteristics of Intel SGX~\cite{sgx}. We then provide a brief overview of consensus protocols. 

\subsection{Intel SGX}
\label{subsec:SGX}

\noindent\textbf{Enclave Execution.} Intel SGX~\cite{sgx} is a set of CPU extensions that are available on Intel CPUs starting from the Skylake microarchitecture~\cite{intel_skylake}, capable of providing hardware-protected TEE (or \textit{enclave}) for generic computations. It enables a host to instantiate one or multiple enclaves simultaneously. An enclave is associated with a CPU-guarded address space which is accessible only by the enclave code; the CPU blocks any non-enclave code's attempt to access the enclave memory. This effectively isolates the enclave from other enclaves concurrently running on the same host, from the OS, and from other user processes. Memory pages can be swapped out of the enclave memory, but they are encrypted using the processor's key prior to leaving the enclave. 

Enclaves cannot directly execute OS-provided services such as I/O. In order to access those services, enclaves have to employ OCalls (calls executed by the enclave code to transfer the control to non-enclave code) and ECalls (API for untrusted applications to transfer control back to the enclave). These ECalls and OCalls constitute the enclave boundary interface, enabling a communication between the enclave code and the untrusted application to service OS-provided functions. Clearly, enclave developer should design such interface with caution, for ECalls exposed to the untrusted application may open up an attack surface to the protected execution environment.

\vspace{2mm}
\noindent\textbf{Attestation.}
Intel SGX allows a validator to verify if a specific enclave has been properly instantiated with the correct code~\cite{sgx_remote_attest} via its attestation mechanisms. These mechanisms also enable the validator and the attesting enclave to establish a secure, authenticated channel via which sensitive data can be communicated. 

If the attestation is carried out between two enclaves instantiated on the same platform (or host), the mechanism in use is referred to as {\em local attestation}. Once the attesting enclave has been initiated, the CPU computes its \textit{measurement} (i.e., the hash of its initial state). Next, the CPU produces a message authentication code (MAC) of such measurement using a key that is accessible only by the validating enclave. The measurement and the MAC are then sent to the validating enclave for verification. Alternatively, if the validator is a remote party, the CPU issues a {\em remote attestation} by signing the measurement with its private key under the Enhance Privacy ID (EPID) scheme~\cite{epid}~\cite{sgx_remote_attest}. The remote party obtaining the attestation then relies on the Intel's Attestation Service (IAS) to verify the signature contained in the attestation~\cite{ias}, and then check the measurement value against a known value. 

\vspace{2mm}
\noindent\textbf{Data sealing.} Enclaves can persist their private state to non-volatile memory via the data sealing mechanism. The enclave first requests the CPU for a enclave-specific key bound to its measurement. It then encrypts its private state using the requested key before storing it on persistent storage. The mechanism ensures that the sealed data can only be accessed by the enclave that sealed it. Nevertheless, enclave recovery using sealed data is susceptible to rollback attacks wherein an attacker (e.g., the malicious OS) provides the enclave with properly sealed but stale data~\cite{rote}.

\subsection{Consensus Protocols}
A rich literature of \textit{consensus protocols} has been devoted to address a variety of fault tolerance problems in distributed systems~\cite{raft, pbft}. Consensus algorithms can be broadly categorized by the failure model that they assume. On the one hand, the {crash failure} model assumes that a faulty process may stop processing any message, and it does not resume~\cite{raft}. On the other hand, the {Byzantine failure} model may allow for arbitrary faults. A process experiencing Byzantine failure may deviate from its expected behaviors in any manner, it may equivocate (i.e., sending contradictory messages to other processes), or it may intentionally delay its activity for any period of time~\cite{pbft}. 

Consensus algorithms aim to achieve \textit{safety} and \textit{liveness} in the presence of failures. Safety requires that non-faulty processes reach an agreement and never return conflicting results for the same request, whereas liveness means that these processes eventually agree on a value. In the following, we give an overview of Raft~\cite{raft}, a consensus protocol that assumes crash failure model and ensures safety regardless of timing. Nonetheless, its liveness depends on timing (e.g., the communication channels between processes are partially synchronous, wherein messages are delivered within an unknown but finite bound). 

\vspace{2mm}
\noindent{\bf Raft Consensus Protocol.}
Raft assumes a system of $n$ deterministic processes (or servers), among which at most $f = \frac{n-1}{2}$ could be faulty. A faulty process fails by crashing. Each process stores a log that contains a series of commands, or events of interest. Raft ensures that logs of non-faulty processes contain the same sequence of commands. 

A process can be in one of the three roles, namely {\it leader, follower} and {\it candidate}. 
Raft divides time into {\it terms} numbered with consecutive integers. In each term, one process is elected as the leader, and all other processes are followers. The leader exchanges periodic heartbeats with all followers to maintain its authority. If a follower does not receive any heartbeat after an \textit{election timeout} period, it assumes that the leader has crashed. The follower then increases its term, changes it role to candidate and begins an election. The candidate becomes the leader if it receives votes from a majority of processes. 
We refer readers to \cite{raft} for further details on the leader election and its election criteria.

During normal operation, the followers are passive, only responding to requests from the leader and candidate. The leader receives all the commands (e.g., requests from the clients), and replicates the commands on the followers. The leader first appends a command as a new entry which is uniquely identified by the leader's \textit{term} and an index to its log. Next, it broadcasts the entry to all of the followers. The followers append the received entry to their logs, and acknowledge the receipt to the leader. Once the leader receives the acknowledgement from a majority of the processes (i.e., $f+1$ or more processes), it \textit{commits} the entry (i.e., execute the command contained in the entry), and all preceding entries in its log. The leader records the highest index it knows to be committed, and includes this index in subsequent messages to the followers, thereby informing them on the committed  entries.

Although Raft is designed for crash failure model, there exists attempt that deploys Raft in a Byzantine setting~\cite{mscoco}. This is achieved through the use of TEEs. More specifically, by running the consensus protocol inside an enclave that offers attested execution, one can restrict adversarial behaviours of the faulty processes, thereby reducing the threat model from Byzantine fault tolerance to crash fault tolerance, to which Raft applies.

\section{The Problem}
\label{sec:problem_statement}
In this section, we first define our system model in Section~\ref{subsec:system_model}. Next, we present the system goals that we seek to achieve in Section~\ref{subsec:system_goal}. We then describe in Section~\ref{subsec:adv_model} the adversary model against which our proposed autonomous membership service is designed. Finally, we detail the challenges of enabling autonomous membership service for enclave processes in Section~\ref{subsec:challenges}.

\subsection{System Model}
\label{subsec:system_model}
We consider a typical outsourced computation model consisting of a cloud service provider (CSP) and its users. The CSP operates a set of machines (or hosts) that are equipped with Intel SGX-enabled processors~\cite{sgx}. A user (i.e., application owner) uploads her code to the cloud, and executes her application on the cloud's infrastructure.  The application involves secrets that the user wishes to keep private from the CSP. This can be achieved by running the application code (or a critical portion of its code) inside an SGX enclave. In particular, the code uploaded to the cloud does not contain any secret.
The SGX architecture then relies on the CSP's hypervisor or OS to instantiate an enclave process using the application code. Subsequently, the enclave process attests itself to the application owner so as to convince the latter that it is properly instantiated. Upon successful attestation, the secret materials are securely provisioned to the enclave. 

The application code can be split into security-critical and non-critical components. The critical component is to be run in an enclave (or enclaves). The CSP may spawn multiple instances of the same application. Each application instance comprises both the critical and non-critical components, and they are bound together. If the critical component is divided into a set of enclaves, we assume that they are also uniquely bound together. In another word, the service provider cannot mix-and-match a component of one instance with that of another instance without breaking the function of the application. Consequently, it suffices to identify an application instance by its enclave (or one of its enclaves). Without loss of generality, we assume throughout the rest of the paper that the application runs entirely in a single enclave, and abuse the language to refer to each instance as a \textit{process}.

\vspace{2mm}
\noindent\textbf{Functional vs. Nonfunctional process.} We assume that the application involves secret materials that are concealed from the CSP or any untrustworthy party, and that such secret is necessary for the application's operation (e.g., encryption/decryption keys). While the CSP has a sole discretion in instantiating a new process using the application code, such a process is initially {\it nonoperational}. Only until it is provisioned with the aforementioned application secret does it become
{\it operational}. Put differently, although the CSP facilitates the commissioning of a new process, it cannot alone complete such a commission.

\vspace{2mm}
\noindent\textbf{Communication.} The processes communicate only by passing messages along a fixed set of channels facilitated by CSP. Each pair of processes is connected via a reliable, authenticated point-to-point communication channel that does not drop any message. These channels, however, are subject to a delivery scheduler controlled by the CSP. This allows the CSP to impose arbitrary delay on any message at any channel of its choice. When a channel is not intervened by a malicious delivery scheduler, it is \textit{synchronous}, in a sense that all messages sent via that channel are delivered within a finite delay $\Delta$ known to all processes. The malicious scheduler can postpone the delivery of any message for an arbitrary duration, causing a \textit{communication fault}, wherein messages communicated via that channel are not delivered within the delay $\Delta$. Finally, we do not assume any global clock, or bounds on relative local clock speed and execution speed of each process. 

\definecolor{applegreen}{rgb}{0.6, 0.8, 0.1}
\begin{figure*}[t]
\begin{subfigure}{.3\textwidth}
\centering  
\resizebox{0.5\textwidth}{!} {
\centering  
\begin{tikzpicture}[every text node part/.style={align=center}]
\node[circle, draw = black, thick, dotted](p1){$p_1$};
\node[circle, draw = black, thick, dotted, below of = p1, node distance = 1cm](p2){$p_2$};
\node[circle, draw = black, thick, below of = p1, node distance = 1.5cm, xshift = 1.5cm , fill =applegreen](p3){$p_3$};
\node[circle, draw = black, thick, above of = p1, node distance = 0.5cm, xshift = 1.5cm, fill =applegreen](p4){$p_4$};
\node[circle, draw = black, thick, right of =p1, node distance = 2.5cm, yshift = -0.5cm, fill =applegreen](p5){$p_5$};
\draw[-, thick] (p4.south east) -- (p5.north west);
\draw[-, thick] (p5.south west) -- (p3.north east);
\draw[-, thick] (p3.north) -- (p4.south);

\end{tikzpicture}
}
\caption{$n = 5, f= 2, i= 0$. $\texttt{Prog}$ is not in anarchy.}
\end{subfigure}
\hfill
\begin{subfigure}{.3\textwidth}
\centering  
\resizebox{0.5\textwidth}{!} {
\centering  
\begin{tikzpicture}[every text node part/.style={align=center}]
\node[circle, draw = black, thick, dashed, fill = pink](p1){$p_1$};
\node[circle, draw = black, thick, dotted, below of = p1, node distance = 1cm](p2){$p_2$};
\node[circle, draw = black, thick, below of = p1, node distance = 1.5cm, xshift = 1.5cm, fill =applegreen](p3){$p_3$};
\node[circle, draw = black, thick, above of = p1, node distance = 0.5cm, xshift = 1.5cm, fill =applegreen](p4){$p_4$};
\node[circle, draw = black, thick, right of =p1, node distance = 2.5cm, yshift = -0.5cm, fill =applegreen](p5){$p_5$};
\draw[-, thick, dashed] (p1.north east) -- node[yshift = 0.2cm, rotate = 10, xshift = -0.1cm] {$> \Delta$} (p4.west);
\draw[-, thick] (p1.south east)-- (p3.west);
\draw[-, dashed, thick] (p1.east) -- node[yshift = 0.2cm, rotate = -10, xshift = -0.1cm] {$> \Delta$}(p5.west);
\draw[-, thick] (p4.south east) -- (p5.north west);
\draw[-, thick] (p5.south west) -- (p3.north east);
\draw[-, thick] (p3.north) -- (p4.south);

\end{tikzpicture}
}
\caption{$n = 5, f= 1, i= 1$. $\texttt{Prog}$ is not in anarchy.}
\end{subfigure}
\hfill
\begin{subfigure}{.3\textwidth}
\centering  
\resizebox{0.5\textwidth}{!} {
\centering  
\begin{tikzpicture}[every text node part/.style={align=center}]
\node[circle, draw = black, thick, dashed, fill = pink](p1){$p_1$};
\node[circle, draw = black, thick, dotted, below of = p1, node distance = 1cm](p2){$p_2$};
\node[circle, draw = black, thick, below of = p1, node distance = 1.5cm, xshift = 1.5cm , fill =applegreen](p3){$p_3$};
\node[circle, draw = black, thick, above of = p1, node distance = 0.5cm, xshift = 1.5cm, fill =applegreen](p4){$p_4$};
\node[circle, draw = black, thick, dashed, fill = pink, right of =p1, node distance = 2.5cm, yshift = -0.5cm](p5){$p_5$};
\draw[-, thick, dashed] (p1.north east) --  node[yshift = 0.2cm, rotate = 10, xshift = -0.1cm] {$> \Delta$}(p4.west);
\draw[-, thick] (p1.south east)-- (p3.west);
\draw[-, dashed, thick] (p1.east) -- node[yshift = 0.2cm, rotate = -10, xshift = -0.1cm] {$> \Delta$}(p5.west);
\draw[-, thick] (p4.south east) -- (p5.north west);
\draw[-, thick, dashed] (p5.south west) --  node[yshift = 0.2cm, rotate = 45, xshift = -0.1cm] {$> \Delta$}(p3.north east);
\draw[-, thick] (p3.north) -- (p4.south);
\end{tikzpicture}
}
\caption{$n = 5, f= 1, i= 2$. $\texttt{Prog}$ is in anarchy.}
\end{subfigure}
\caption{Illustrations of faulty, isolated and active processes. They are depicted in dotted, dashed and solid circles, respectively.}
\label{fig:fault_example}
\end{figure*}
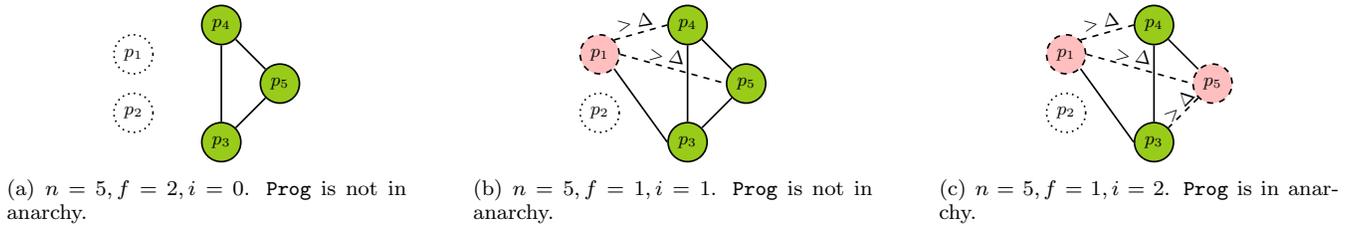

\vspace{2mm}
\noindent\textbf{Faults.} We call a process \textit{faulty} if it has crashed or halted. A process is \textit{correct}, otherwise. Under the communication model described earlier, correct processes may be \textit{isolated} from other correct processes.  
Let us consider an application $\texttt{Prog}$. We denote by $n_s(\texttt{Prog})$ the number of $\texttt{Prog}$'s existing processes at a given moment $s$. We deem a process $p$ isolated if $p$ cannot communicate with a quorum of $\lfloor \frac{n_s(\texttt{Prog})}{2} \rfloor$ processes in a synchronous manner. 
We call a process \textit{active} if it is correct and not isolated.

Let $f_s(\texttt{Prog})$ and $i_s(\texttt{Prog})$  be the number of faulty and isolated processes among the 
$n_s(\texttt{Prog})$ processes, respectively. 
We deem the application $\texttt{Prog}$ to be in \textit{anarchy} at the given moment $s$  if:

\begin{equation}
\label{eq:anarchy}
f_s(\texttt{Prog}) + i_s(\texttt{Prog}) > \lfloor \frac{n_s(\texttt{Prog})-1}{2} \rfloor
\end{equation}

Hereafter, when it is clear from the context, we omit $s$ and $\texttt{Prog}$ from the notations. Figure~\ref{fig:fault_example} depicts different faulty scenarios $\texttt{Prog}$'s processes may experience, and state where $\texttt{Prog}$ is in anarchy.


\subsection{System Goals}
\label{subsec:system_goal}
To allow for elasticity, the CSP needs to commission or decommission processes dynamically overtime. Nonetheless, this may expose the system to forking attacks, as described earlier in Section~\ref{sec:intro}. To address this problem, we seek an autonomous membership service that protects the application against such forking attacks. The membership service polices the commissioning of new processes, and excommunication of faulty or isolated processes from the system, subject to the SLA concerning the upkeeping of the application. More specifically, while the CSP can instantiate a new process at will,
such a process remains nonoperational until its request to join the existing group of processes is approved by the membership service. In addition, the membership service actively monitors the status of each member process, and excommunicates a process that it suspects of being faulty or isolated from the group. 

To render the membership service autonomous, our approach is to augment each process of an application \texttt{Prog} with a membership list that keeps track of \texttt{Prog}'s active processes, and ensure that the membership lists of all active processes converge without relying on manual
intervention of the application owner, or a trusted third party:

\begin{itemize}
\item If a process becomes faulty, it is required that the faulty process is eventually suspected of having halted, and excommunicated from the membership list of all active processes. Communication from the excommunicated process are discarded. 

\item A newly instantiated process must join the system via the membership service before it is rendered functional. The membership service ensures that this new process is added to the membership list of all existing active processes, and the former is brought up-to-date with the current system membership list. 

\item Except for the instantiation of the first process, the application owner does not involve in operation of the membership service. All subsequent process attestation and secret provisioning are handled by active processes in the system. 
\end{itemize}

\subsection{Adversary Model}
\label{subsec:adv_model}
We consider a powerful adversary that possesses all capabilities of the CSP. In practice, such adversary can be an errant insider having complete access to the cloud infrastructure, or an attacker exploiting vulnerabilities in the cloud's software stack. The adversary is interested in conducting a forking attack on the victim application, attempting to launch an arbitrarily large number of enclave instances in violation of the SLA, or to isolate the concurrently running instances from each other. Nonetheless, we assume that the adversary does not seek to undermine the application's availability (e.g., by decommissioning all of its instances). 

The adversary has complete control over the cloud's OS and/or hypervisor. It can schedule or corrupt all non-enclave processes, and restart enclave instances at will. Moreover, it can control all network communications between the enclave instances regardless of whether they are instantiated on the same or across different hosts. More specifically, the adversary can read, modify, reorder, delay or even drop the messages sent by the enclaves arbitrarily. 

 Nonetheless, it cannot compromise protection mechanisms of the SGX processors. That is, it can neither access the enclave runtime memory nor learn the application secrets that are protected by the enclave execution, and does not have any access to the processor's private keys that are used for attestation and data sealing functions. Finally, we assume that the adversary is computationally bounded, and cannot break standard cryptographic assumptions.

We assume the application code running inside the enclave is secure. We do not consider side-channel attacks against the enclave execution~\cite{controlled_channel} and DoS attacks against the system.

\subsection{Challenges}
\label{subsec:challenges}
The first challenge in enabling autonomous membership service stems from the need of injecting application secret to the newly instantiated process. 
Since the untrusted OS is in charge of enclave instantiating, any secret material necessitated for the operations of the enclave process cannot be contained in the enclave code, but should rather be provisioned to the enclave only after it has been properly instantiated. 
In a typical pipeline of enclave instantiating, this secret provisioning is realised via the remote attestation procedure carried out by the application owner. 
Nonetheless, it is unreasonable to assume the involvement of the application owner in the instantiating
of every enclave process, especially when processes may be commissioned dynamically to meet the elasticity requirement of the application.
Ideally, the application owner only needs to involve in the instantiating of the first enclave instance during the application provisioning process, and does not have to take part in the upkeeping of her application.

The second challenge lies in ensuring the security of the membership service itself, especially against forking attack~\cite{forking_attacks} or logical partition. To further elaborate on this challenge, we describe in the following two straw-man solutions and point out their weaknesses. The first solution assumes a trusted \textit{directory server} keeping track of all running processes of the application,
thereby offering the membership service. The second solution attempts to implement the \textit{directory server} as a replicated state machine so as to avoid single-point-of-failure. 

\vspace{2mm}
\noindent\textbf{Directory server.} A trusted directory server is in charge of managing and attesting the correct instantiating of application processes, and keeping track of all such running processes. More specifically, the user uploads the expected measurement of her application enclave process (i.e., the hash of its initial state) along with its secret to the directory server via a secure channel. The directory server then performs a remote attestation with the newly instantiated enclave process on behalf of the user, and provides it with the application secret upon successful attestation. It maintains a list of concurrently running processes, and refuses the instantiating of the new process if it violates the SLA. 

The directory server determines the aliveness of a process in its process list by listening to a \texttt{heartbeat} message sent by that process, to which it responds with an acknowledgement \texttt{ack}. If it does not receive the \texttt{heartbeat} from the process after a certain time-out period, it would assume that such process has crashed, removing such process from its process list.
It is important to note that the excommunication of a process may be due to communication fault, rather than the said process having halted (i.e., unjust excommunication).
In order to keep the logic of the directory server simple, we can enforce the unjust excommunication by forcing the excommunicated process to halt. This can be achieved by programming a process to \textit{commit suicide} if it does not receive an \texttt{ack} from the directory server after a configurable \texttt{time-to-live} period, and to restart its \texttt{time-to-live} timer every time it receives an \texttt{ack}. Put differently, the \texttt{ack} from the directory server acts as a \textit{lease} that allows the receiving process to continue operating for a fixed period of time until it expires~\cite{leases}.

This approach, nevertheless, is not desired, for the directory server becomes a single-point-of-failure, or target for attacks. For example, if the directory server is implemented using TEE, it is subject to rollback attack~\cite{rote} in which the adversary may deceive it to resume operations using stale process list, violating the integrity and consistency of the directory server's operations. 

\vspace{2mm}
\noindent\textbf{Replicated directory service.}
One approach to sidestep the aforementioned single-point-of-failure issue is by replicating the directory server. In particular, the directory service now comprises of a set of replicas that leverage fault-tolerant protocol to implement a replicated state machine, ensuring consistency of the service's operation despite individual replica failures~\cite{replicatee}. Nonetheless, it is challenging to protect these replicas against forking attack~\cite{forking_attacks}, wherein the adversary manage to splits the directory service replicas into two ``cliques'', each of which is unaware of the other. Such attack allows the adversary to run twice as many number of processes as what would otherwise be allowed by the SLA.  Addressing this issue, then, requires a membership service offered to the directory service replicas themselves.

\section{Our Design}
\label{sec:design}

In this section, we present our system \codename~(\underline{A}utonomous \underline{ME}mbership \underline{S}ervice). 
We first give an overview of \codename's design. We later discuss in details the main protocols that comprise the autonomous membership service. 
We remark that although our discussion assumes processes are provisioned as SGX enclaves, the technique underlying \codename\ is applicable to other Trusted Execution Environment platforms that support similar features to SGX enclaves.

\subsection{Overview}
\label{subsec:overview}
To enable the autonomous membership service for an application \texttt{Prog}, \codename\ augments \texttt{Prog}'s processes to equip them with the following parameters:

\vspace{1mm}
\begin{itemize}
\itemsep1mm 

\item \texttt{peerList}: a list that records all active processes of \texttt{Prog}. 

\item term number: a value that represents the current term (which we shall define in the following). The term number increases monotonically. 

\item replicated log: a log that records messages the process exchange during the operation of the membership service. A log entry is indexed using the term number of the term at which it is added, together with sequence number that increases monotonically over time across terms.

\item commitIndex: the index of the latest entry in the replicated log that is known to be committed. This value increases monotonically over time.

\item heartbeat timeout \texttt{T}: a randomized timeout used in \codename's \texttt{heartbeat} mechanism, unique for each process.

\item candidature timeout: the timeout used during the leader election. This timeout is the same for all processes.
\end{itemize}

While a process can always derive \texttt{peerList} from its replicated log, and one may argue that they can be merged into a single data structure, we find that treating them as two separate entities keep our exposition clearer.



\codename\ requires the involvement of the application owner during the application bootstrap so as to attest the correct instantiating of the first process, and to provide it with the application secret (Section~\ref{subsec:bootstrap}). 
Once the first instance is fully operational, the application owner may go offline until the time when she wishes to shutdown the application. 
All subsequent process instantiating (or dismantling) over the lifetime of the application will be policed by \codename. To this end, \texttt{Prog} needs to incorporate in its implementation \codename's protocols for managing the processes' \texttt{peerList}s and ensuring that they converge. We describe these protocols in the remaining of this section, deferring discussion on the implementation details to Section~\ref{sec:implementation}.

\vspace{2mm}
\noindent\textbf{Workflow.}
Let $n$ denote the number of processes of \texttt{Prog} at a given time. \codename\ first elects one process to be the {\it leader}, while the other $[n-1]$ processes are {\it followers}. 
A leader is associated with a \textit{term}, which is a time period of arbitrary length in which the leader retains its authority. 
The leader relies on a \texttt{heartbeat} mechanism to monitor the status of the processes in the system, and to maintain its authority over them (Section~\ref{subsec:heartbeat}). In normal operation (i.e., the leader is active), the followers need not exchange \texttt{heartbeat} messages among themselves, but simply respond to messages from the leader. 

The commission of a new process requires an ``approval'' from the membership service
via a \newinstance\ protocol orchestrated by the leader. More specifically, the leader attests the new process, and checks against the SLA if its commission is eligible (e.g., if the quota of concurrently running processes has not been consumed). 
The leader then provisions the application secret to the new process, rendering it functional, and informs all followers to update their \texttt{peerList}s accordingly (Section~\ref{subsec:new_instance}). If the leader suspects a follower process $\texttt{p}_{\texttt{h}}$ of having halted (e.g., it did not receive a response from $\texttt{p}_{\texttt{h}}$ for the \texttt{heartbeat} message it has sent to $\texttt{p}_{\texttt{h}}$ after a configurable heartbeat timeout $\texttt{T}_\texttt{h}$ which is unique to $\texttt{p}_{\texttt{h}}$), it drives the other followers to expel $\texttt{p}_{\texttt{h}}$ from their \texttt{peerList} via the \failedinstance\ protocol (Section~\ref{subsec:failed_instance}).
Finally, when the user decides to shutdown the application, she sends the shutdown request to the leader, who then coordinates the termination of all active processes via a \shutdown\ protocol, at the end of which they all cease to operate (Section~\ref{subsec:shutdown}). 

It is important to note that \codename\ admits unjust excommunication of a active process whose connection with the leader is severed. To prevent the excommunicated process from disrupting the membership service, its future communication (i.e., after the point of excommunication) will be discarded, and the process will eventually halt.

The workflow described above has not accounted for leader failure. If the leader halts or becomes isolated, followers will eventually detect such failure, and engage in a leader election protocol to elect a new leader. In particular, if a follower does not receive a \texttt{heartbeat} message from the leader after its heartbeat  timeout, it will assume that the current leader is no longer viable, and attempt to claim the leadership for itself (Section~\ref{subsec:heartbeat}). When a new leader is elected, the new term starts.  During the period where a process attempting to claim the leadership, it is referred to as a \textit{candidate}. More than one candidates may compete for the leadership at the same time. The leader election guarantees that only one candidate succeeds, and all active processes in the system recognize the new leader as legitimate. 


\subsection{Application Bootstrap}
\label{subsec:bootstrap}
In \codename, application bootstrap refers to the instantiating of the very first process of the application. This procedure is initiated by the application owner, who instructs the cloud service provider to instantiate an enclave process using the application code. Since the cloud's OS/hypervisor is not trusted, the instantiating has to be attested by the application owner, and application secret is only provisioned to the enclave after a successful attestation. 

\codename\ benefits directly from Intel SGX's remote attestation protocol~\cite{sgx_remote_attest} in implementing the above mentioned attestation and secret provisioning. We highlight key characteristics of the protocol, and refer the readers to~\cite{sgx_remote_attest} for further details. The SGX architecture features a special enclave, called {\em Quoting Enclave}, to facilitate the remote attestation. After being instantiated, the attesting enclave invokes the SGX's \texttt{EREPORT} instruction and specifies the Quoting Enclave as its target. The \texttt{EREPORT} creates a structure, called \texttt{report}, that contains all necessary information to verify the correctness of the attesting enclave's instantiating, and produces a message authentication code (MAC) tag for the report. The MAC is computed using a ``report key'' that is associated only with the attesting enclave and its intended target (i.e., Quoting Enclave in this case). The Quoting Enclave verifies the \texttt{report}, and subsequently replaces its MAC with a signature signed using a CPU's private key under the EPID group signature scheme~\cite{epid}, producing a remote attestation \texttt{quote}\footnote{EPID allows a signer to sign objects with his private key without uniquely identifying the signer or linking his different signatures. This is achieved by assigning each signer to a ``group" such that the group's public key can be used to verify all signatures produced by the group member.}. As Intel does not publish the group public key associated with the CPUs' private key, the verifier has to rely on the Intel Attestation Service to validate the remote attestation \texttt{quote}. 

It is worth noting that SGX remote attestation mechanism allows the verifier to include a challenge in her attestation requests, and the attesting enclave to include a manifest that contains response to the verifier's challenge and an ephemerally generated public key to the signed struture \texttt{report}, which is then carried over to the \texttt{quote}~\cite{sgx_remote_attest}. The challenge and response ensures freshness of the \texttt{quote}, avoiding replay attack by the malicious OS. The ephemeral public key allows the verifier to securely provision application secrets to the enclave. Consequently, upon successful application bootstrap, the first application process becomes fully functional, containing the secrets necessitated for the application's operation. 


\subsection{\heartbeatexchange\ and Leader Election}
\label{subsec:heartbeat}
During normal operation, the leader uses \texttt{heartbeat} mechanism to maintain its leadership and to monitor status of follower processes. A \texttt{heartbeat} is essentially a message sent by the leader, and acknowledged with a corresponding \texttt{ack} message by a follower, via an authenticated channel. A 
\texttt{heartbeat} always contains the term number and commitIndex of the leader (the latter is the index of the latest entry in the leader's replicated log that has been committed). The follower relies on the leader's commitIndex to detect if its replicated log and state are up-to-date with those of the leader. If a follower finds that its log misses some entries, or its state is out of date, it fetches the missing log entries and commit them, bringing its state up-to-date. The \texttt{heartbeat} can also contain information regarding the commission (excommunication) of a new (existing) follower process, which we shall elaborate in the following subsections. 

When a new process becomes functional, it initially assumes the role of a follower. The follower process $\texttt{p}_{\texttt{i}}$ and the leader process $\texttt{p}_{\texttt{L}}$ establish a randomized timeout $\texttt{T}_{\texttt{i}}$, which is unique to each follower process, chosen from a fixed interval. 
Via the \texttt{heartbeat} mechanism, $\texttt{p}_{\texttt{L}}$ can identify if it has become isolated, and halts if it is so. On the other hand, if $\texttt{p}_{\texttt{L}}$ is active but cannot receive an \texttt{ack} message from $\texttt{p}_{\texttt{i}}$ over a period of $\texttt{T}_{\texttt{i}}$, it suspects the latter of having halted, and triggers the \failedinstance\ protocol to inform all other followers of $\texttt{p}_{\texttt{i}}$ failure. Alternatively, if the follower $\texttt{p}_{\texttt{i}}$ is active but cannot receive a \texttt{heartbeat} from the leader over a period of $\texttt{T}_{\texttt{i}}$, it changes its role to \textit{candidate}, increases its term number and broadcasts a $\texttt{RequestVote}$ message to all existing processes in its $\texttt{peerList}$ to claim the leadership. 

The $\texttt{RequestVote}$ message from $\texttt{p}_{\texttt{i}}$ takes a form of $\langle \texttt{p}_{\texttt{i}}, \texttt{newTerm}, \texttt{lastLogIndex} \rangle$, where \texttt{newTerm} is its current term number (which has just been increased), and \texttt{lastLogIndex} is the  index of the last entry in its replicated log. Upon receiving $\texttt{RequestVote}$ from $\texttt{p}_{\texttt{i}}$, a receiver process $\texttt{p}_{\texttt{r}}$ acts as follows:

\begin{itemize}
\item If $\texttt{p}_{\texttt{i}}$ is not in $\texttt{p}_{\texttt{r}}$'s $\texttt{peerList}$, it ignores $\texttt{p}_{\texttt{i}}$'s $\texttt{RequestVote}$.

\item If $\texttt{p}_{\texttt{i}}$ appears in its $\texttt{peerList}$, but $\texttt{p}_{\texttt{r}}$ has recently received a message from the leader $\texttt{p}_{\texttt{L}}$ (i.e., its timeout $\texttt{T}_{\texttt{r}}$ has not expired), $\texttt{p}_{\texttt{r}}$ queues $\texttt{p}_{\texttt{i}}$'s $\texttt{RequestVote}$ if it has not queued any other $\texttt{RequestVote}$. 
If $\texttt{p}_{\texttt{r}}$ has queued a $\texttt{RequestVote}$ from another candidate $\texttt{p}_{\texttt{c'}}$, $\texttt{p}_{\texttt{r}}$ determines the dominating candidate among $\texttt{p}_{\texttt{i}}$ and $\texttt{p}_{\texttt{c'}}$\footnote{We say a candidate $\texttt{p}_{\texttt{c}}$ dominates (or outweights) another candidate $\texttt{p}_{\texttt{c'}}$ if it has a more up-to-date log, which is determined by the index of last entries in the logs~\cite{raft}.}, 
keeping only the $\texttt{RequestVote}$ of the dominating candidate, and informing the dominated candidate on the existence of the dominating one. 
Once its \texttt{heartbeat} timeout $\texttt{T}_{\texttt{r}}$ expires, $\texttt{p}_{\texttt{r}}$ check if it is dominated by the candidate whose $\texttt{RequestVote}$ is currently kept in its queue. If so,
it grants its vote to the candidate; otherwise it discards such a $\texttt{RequestVote}$ and broadcasts its own. 
On the other hand, if $\texttt{p}_{\texttt{r}}$ receives the next \texttt{hearbeat} from $\texttt{p}_{\texttt{L}}$ before $\texttt{T}_{\texttt{r}}$ expires, it resets $\texttt{T}_{\texttt{r}}$ and discards any $\texttt{RequestVote}$ in its queue.

\item If $\texttt{p}_{\texttt{i}}$ is in its $\texttt{peerList}$, but it has voted for another candidate (potentially itself) that is not dominated by $\texttt{p}_{\texttt{i}}$, $\texttt{p}_{\texttt{r}}$ informs $\texttt{p}_{\texttt{i}}$ on such candidate.

\end{itemize}

If $\texttt{p}_{\texttt{i}}$ collects a quorum of votes from a \textit{majority} of the processes in its \texttt{peerList} for its $\texttt{RequestVote}$ before its heartbeat timeout expires, it assumes the role of the leader. It then drives the processes to expel $\texttt{p}_{\texttt{L}}$ from their \texttt{peerList}.
While waiting for the votes, if $\texttt{p}_{\texttt{i}}$ is informed of another candidate that outweighs itself, $\texttt{p}_{\texttt{i}}$ resets its heartbeat timeout and waits for the other candidate to pronounce its leadership. 
When $\texttt{p}_{\texttt{i}}$ is informed of a new leader whose term is as large as its current term via a \texttt{heartbeat} from the latter, it acknowledges the leader's \texttt{heartbeat}, and returns its role to follower.
If $\texttt{p}_{\texttt{i}}$ fails to collect a quorum of responses from a majority of the processes, or a \texttt{heartbeat} from the new leader within its heartbeat timeout, it increases its term and triggers a new election with a new $\texttt{RequestVote}$.
Finally, if $\texttt{p}_{\texttt{i}}$ is still assuming the candidate role (i.e., it can neither claim the leadership nor return to the follower role) when its candidature timeout expires, it commits suicide.

\vspace{2mm}
\noindent\textit{Security Arguments:} The leader election in \codename\ follows Raft's leader election~\cite{raft} to certain extent. The key difference here is that a candidate $\texttt{p}_{\texttt{i}}$ shall halt (by committing suicide) if it cannot collect a quorum of responses or a \texttt{heartbeat} from the new leader within the candidature timeout. Here, we reason about the correctness of this design when the system is not in anarchy.

The fact that $\texttt{p}_{\texttt{i}}$ issuing its $\texttt{RequestVote}$ implies either the leader $\texttt{p}_{\texttt{L}}$ has become faulty or isolated, or the connection between $\texttt{p}_{\texttt{i}}$ and the active $\texttt{p}_{\texttt{L}}$ has been severed. 

If $\texttt{p}_{\texttt{L}}$ is isolated, it will detect its isolation by itself and halts, thereby becoming faulty. When $\texttt{p}_{\texttt{L}}$ is indeed faulty\footnote{In this case, no new process can be commissioned, and no process failures can be detected during the period when the system is in the leader election stage.}, other processes will either grant $\texttt{p}_{\texttt{i}}$ their vote or inform it on the dominating candidate. If $\texttt{p}_{\texttt{i}}$ is active (i.e., it is not isolated), it will have received enough responses to either claim the leadership, and thereby orchestrating the removal of $\texttt{p}_{\texttt{L}}$, or to wait to be contacted by the new leader. 
On the other hand, if $\texttt{p}_{\texttt{L}}$ is active, it would have simultaneously driven other processes to expel $\texttt{p}_{\texttt{i}}$ from their \texttt{peerList}, preventing them from responding to $\texttt{p}_{\texttt{i}}$'s $\texttt{RequestVote}$. This, in turn, prevents $\texttt{p}_{\texttt{i}}$ from being able to collect a quorum of responses for its $\texttt{RequestVote}$, which subsequently leads to $\texttt{p}_{\texttt{i}}$'s committing suicide. 

\vspace{2mm}
\noindent\textit{Remark.} While our design admits a potential unjust excommunication of an active process whose connection to the active leader is severed, this is necessary to sidestep the difficulty of solving membership problem in an asynchronous environment when faults may be present~\cite{flp, birman_membership}. By requiring the active yet excommunicated process to commit suicide, \codename\ preserves the consistency of the leader's expelling decision and safety of the membership service.

\subsection{\newinstance\ Protocol}
\label{subsec:new_instance}
After the application has been bootstrapped and as long as it is not in anarchy, \codename\ enables the application to commission new processes in an autonomous manner, without relying on manual intervention of the application owner, or a trusted third party. This section describes how \codename\ commissions a new process $\texttt{p}_{\texttt{j}}$ and ensures that the \texttt{peerList}s of all active processes are updated accordingly and that they converge. 
Let $n$ be the number of existing processes at the beginning of $\texttt{p}_{\texttt{j}}$'s commission, and assume that the SLA allows $n>2$.

\vspace{2mm}
\noindent\textbf{Case} $n=1$. The only running and functional process at the beginning of $\texttt{p}_{\texttt{j}}$'s commission is the leader $\texttt{p}_{\texttt{L}}$. The interaction between the two processes happen in two phases. In the first phase, $\texttt{p}_{\texttt{j}}$ contacts $\texttt{p}_{\texttt{L}}$ for attestation. The results of this attestation are two-fold. First, $\texttt{p}_{\texttt{L}}$ is convinced that $\texttt{p}_{\texttt{j}}$ is correctly instantiated. Second, the two processes establish a secure and authenticated communication channel. In the second phase, $\texttt{p}_{\texttt{L}}$ updates its \texttt{peerList} to include $\texttt{p}_{\texttt{j}}$, provisions the application secret as well as its current \texttt{peerList} to $\texttt{p}_{\texttt{j}}$ via the secure channel and establishes a heartbeat timeout $\texttt{T}_\texttt{j}$. Likewise, $\texttt{p}_{\texttt{j}}$'s \texttt{peerList} now contains itself and $\texttt{p}_{\texttt{L}}$. At this point, $\texttt{p}_{\texttt{j}}$ becomes fully functional, and assumes a role of a follower.

\vspace{2mm}
\noindent\textit{Security Arguments:}  The interaction between $\texttt{p}_{\texttt{L}}$ and $\texttt{p}_{\texttt{j}}$ described earlier follows the workflow of a 2-phase-commit (2PC) protocol~\cite{2PC}, with  $\texttt{p}_{\texttt{L}}$ assuming the role of the coordinator. 
The attestation in the first phase acts as a prepare phase in the 2PC protocol, whereas the provisioning of the application secret to $\texttt{p}_{\texttt{j}}$ and updating of \texttt{peerList}s mirror the commit phase of the 2PC protocol. Provided that the application is not in anarchy (i.e., $\texttt{p}_{\texttt{L}}$ remains active), our \newinstance\ protocol does not suffer from blocking, and the \texttt{peerList}s of $\texttt{p}_{\texttt{L}}$ and $\texttt{p}_{\texttt{j}}$ converge.

\vspace{2mm}
\noindent\textbf{Case} $n=2$.  Let us call the two existing processes  $\texttt{p}_{\texttt{L}}$ and $\texttt{p}_{\texttt{2}}$, with $\texttt{p}_{\texttt{L}}$ being the leader. The commissioning of $\texttt{p}_{\texttt{j}}$ happens in two phase. In the first phase, $\texttt{p}_{\texttt{L}}$ engages in a remote attestation with $\texttt{p}_{\texttt{j}}$, and at the same time queries $\texttt{p}_{\texttt{2}}$ (this query is piggybacked on a \texttt{heartbeat} sent to $\texttt{p}_{\texttt{2}}$) if it is ready to update its \texttt{peerList} to contain $\texttt{p}_{\texttt{j}}$. 
In the second phase, if the attestation checks out, and $\texttt{p}_{\texttt{2}}$ responds with a ``yes'' vote (the vote is piggybacked on the \texttt{ack} message corresponding to the \texttt{heartbeat} carrying the query), $\texttt{p}_{\texttt{L}}$ updates its \texttt{peerList} to include $\texttt{p}_{\texttt{j}}$, transfers the application secret as well as the current \texttt{peerList} to $\texttt{p}_{\texttt{j}}$, and establishes a heartbeat timeout $\texttt{T}_\texttt{j}$ with $\texttt{p}_{\texttt{j}}$. $\texttt{p}_{\texttt{L}}$ also informs $\texttt{p}_{\texttt{2}}$ to update its \texttt{peerList} accordingly. 
$\texttt{p}_{\texttt{j}}$ now becomes fully functional, and assume a role of a follower.

\vspace{2mm}
\noindent\textit{Security Arguments:}  Similar to the case where $n=1$, the interaction between $\texttt{p}_{\texttt{L}}$, $\texttt{p}_{\texttt{2}}$, and $\texttt{p}_{\texttt{j}}$ described above follows the workflow of a 2PC protocol~\cite{2PC}, with  $\texttt{p}_{\texttt{L}}$ assuming the role of the coordinator. Provided that the application is not in anarchy (i.e., both existing processes are active), our \newinstance\ protocol does not suffer from blocking, and the \texttt{peerList}s of all processes  converge at the end of $\texttt{p}_{\texttt{j}}$'s commission.

\vspace{2mm}
\noindent\textbf{Case} $n>2$. The leader $\texttt{p}_{\texttt{L}}$ first checks if the commission of $\texttt{p}_{\texttt{j}}$ is eligible according to the SLA. If so, it engages in a remote attestation with $\texttt{p}_{\texttt{j}}$. 
Next, $\texttt{p}_{\texttt{L}}$ follows Raft~\cite{raft} to informing all other followers about the commission of $\texttt{p}_{\texttt{j}}$:

\begin{enumerate}
\item $\texttt{p}_{\texttt{L}}$ puts an entry $\langle \texttt{add}, \texttt{p}_{\texttt{j}} \rangle$ to its log. 

\item $\texttt{p}_{\texttt{L}}$ broadcasts $\langle \texttt{add}, \texttt{p}_{\texttt{j}} \rangle$ to all the followers. 

\item Upon receiving $\langle \texttt{add}, \texttt{p}_{\texttt{j}} \rangle$, a follower $\texttt{p}_{\texttt{i}}$ acknowledges the receipt by sending $\langle \texttt{ack}_{\texttt{i}}, \texttt{add}, \texttt{p}_{\texttt{j}} \rangle$ to $\texttt{p}_{\texttt{L}}$, and puts $\langle \texttt{add}, \texttt{p}_{\texttt{j}} \rangle$ to its log. 

\item Once $\texttt{p}_{\texttt{L}}$ confirms that $\langle \texttt{add}, \texttt{p}_{\texttt{j}} \rangle$ has been replicated on a majority of the processes, its adds $\texttt{p}_{\texttt{j}}$ to its \texttt{peerList},  establishes a heartbeat timeout $\texttt{T}_\texttt{j}$ with $\texttt{p}_{\texttt{j}}$, and transfers to $\texttt{p}_{\texttt{j}}$ the application secret and its \texttt{peerList}. In another word, $\texttt{p}_{\texttt{L}}$ ``commits'' the entry $\langle \texttt{add}, \texttt{p}_{\texttt{j}} \rangle$.

\item $\texttt{p}_{\texttt{L}}$ announces the commit of $\langle \texttt{add}, \texttt{p}_{\texttt{j}} \rangle$ in the next message it exchanges with the followers. Once the followers see the commit, they add $\texttt{p}_{\texttt{j}}$ to their \texttt{peerList}. 
\end{enumerate}

\vspace{2mm}
\noindent\textit{Security Arguments:} Provided that the application is not in anarchy, the \texttt{peerList}s of all active processes converge. This follows directly from the security guarantees of Raft~\cite{raft}, which ensures protocol safety as long as any majority of the processes are active (i.e., they are operational and can communicate with each other without communication fault).

\subsection{\failedinstance\ Protocol}
\label{subsec:failed_instance}
When the leader $\texttt{p}_{\texttt{L}}$ suspects that a process $\texttt{p}_{\texttt{h}}$ has become faulty (e.g., it does not receive a response for the \texttt{heartbeat} sent to $\texttt{p}_{\texttt{h}}$ after the timeout $\texttt{T}_{\texttt{h}}$ bound to $\texttt{p}_{\texttt{h}}$),
it drives all other followers to expel $\texttt{p}_{\texttt{h}}$ from their \texttt{peerList}. 

\begin{enumerate}
\item $\texttt{p}_{\texttt{L}}$ puts an entry $\langle \texttt{Expel}, \texttt{p}_{\texttt{h}} \rangle$ to its log. 

\item $\texttt{p}_{\texttt{L}}$ broadcasts $\langle \texttt{Expel}, \texttt{p}_{\texttt{h}} \rangle$ to all the followers. 

\item A follower $\texttt{p}_{\texttt{i}}$ acknowledges the receipt by sending $\langle \texttt{ack}_{\texttt{i}}, \texttt{Expel}, \texttt{p}_{\texttt{h}} \rangle$ to $\texttt{p}_{\texttt{L}}$, and puts $\langle \texttt{Expel}, \texttt{p}_{\texttt{h}} \rangle$ to its log.

\item Once $\texttt{p}_{\texttt{L}}$ confirms that $\langle \texttt{Expel}, \texttt{p}_{\texttt{h}} \rangle$ has been replicated on a majority of the processes (\textit{excluding} $\texttt{p}_{\texttt{h}}$), its expels $\texttt{p}_{\texttt{h}}$ from its \texttt{peerList}. In another word, $\texttt{p}_{\texttt{L}}$ commits the entry $\langle \texttt{Expel}, \texttt{p}_{\texttt{h}} \rangle$.

\item $\texttt{p}_{\texttt{L}}$ announces the commit of $\langle \texttt{Expel}, \texttt{p}_{\texttt{h}} \rangle$ in the next message it exchanges with the followers. Once the followers see the commit, they expel $\texttt{p}_{\texttt{h}}$ from their \texttt{peerList}. 
\end{enumerate}

\vspace{2mm}
\noindent\textit{Security Arguments:} Since the \failedinstance\ protocol is essentially an instance of Raft~\cite{raft}, it follows from the security guarantees of Raft that the \texttt{peerList}s of all processes at the end of the \failedinstance\ protocol converge, so long as the application is not in anarchy.
It is worth noting that \codename\ admits unjust excommunication of a correct but potentially isolated process or an active process whose connection to the leader $\texttt{p}_{\texttt{L}}$ is severed. 
To prevent the excommunicated process from disrupting the membership service,
all its future communication will be discarded. This is enforced by having other processes not
responding to message from a process that is not in their \texttt{peerList}. 
This would also lead the unjustly excommunicated process to halting. In particular, the unjustly excommunicated process will trigger a leader election, for it has not received any \texttt{heartbeat}
from the leader $\texttt{p}_{\texttt{L}}$. Nonetheless, it cannot obtain the required quorum of responses from other processes for its $\texttt{RequestVote}$, thereby committing suicide as mentioned in Section~\ref{subsec:heartbeat}.

\subsection{\shutdown\ Protocol}
\label{subsec:shutdown}

When the application owner wishes to shutdown all processes of the application, she sends the shutdown request to the leader $\texttt{p}_{\texttt{L}}$, who then coordinates the termination of all active processes as follows:

\begin{enumerate}
\item $\texttt{p}_{\texttt{L}}$ puts an entry $\langle \texttt{preShutdown} \rangle$ to its log. 
It seals the application data, if any, to persistent storage using SGX's data sealing mechanism.

\item $\texttt{p}_{\texttt{L}}$ broadcasts $\langle \texttt{preShutdown} \rangle$ to all the followers. 

\item A follower $\texttt{p}_{\texttt{i}}$ puts the received $\langle \texttt{preShutdown} \rangle$ to its log. Next, it seals the application data, if any, to persistent storage, and responds to the leader by sending $\langle \texttt{ack}_{\texttt{i}}, \texttt{preShutdown} \rangle$.

\item Once $\texttt{p}_{\texttt{L}}$ confirms that $\langle \texttt{preShutdown} \rangle$ has been replicated on a majority of the processes, it broadcasts $\langle \texttt{commitShutdown} \rangle$ to all the followers, thereby announcing the committing of the shutdown request. Finally, its informs the application owner that shutdown request has been served, and halts.

\item Once the followers see the $\langle \texttt{commitShutdown} \rangle$, they halt.

\end{enumerate}
\vspace{2mm}
\noindent\textit{Security Arguments:}
Similar to the \failedinstance\ protocol, the \shutdown\ protocol is an instance of Raft~\cite{raft}. It follows from the security guarantees of Raft that so long as the application is not in anarchy, all active processes shall respond to the shutdown request, and halt. 
Even if the application is in anarchy, as long as the leader commits suicide, all other processes will eventually halt (as we discuss in the next subsection), thereby also bringing the shutdown request into effect.


\subsection{Anarchy}
The application is in anarchy if the total number of faulty and isolated processes (denoted by $f$ and $i$ respectively) exceeds $ \lfloor \frac{n-1}{2} \rfloor$.
When the application falls into anarchy, the status of the current leader $\texttt{p}_{\texttt{L}}$ becomes critical. If $\texttt{p}_{\texttt{L}}$ remains active, the application may still be operational. 
On the other hand, if it is either faulty or isolated, all existing processes will eventually halt, thereby shutdown the entire application. We remark that this only affects availability of the application, not the security of the membership service. More specifically, even if the application is in anarchy, the \texttt{peerList} of active processes do not diverge. In the following, we examine the situations where the application is in anarchy and how the status of the leader $\texttt{p}_{\texttt{L}}$ affects the operation of the application. 

\vspace{2mm}
\noindent{\textbf{Case}} $( f > \lfloor  \frac{n-1}{2} \rfloor  )$:
If the leader $\texttt{p}_{\texttt{L}}$ is among the $f$ faulty processes, the remaining correct processes will trigger leader election. Nonetheless, since $f > \lfloor \frac{n-1}{2} \rfloor $, none of them will be able to obtain a sufficient quorum of responses for their $\texttt{RequestVote}$. Consequently, all correct process shall eventually halt by committing suicide, which effectively shuts the application down in its entirety.

If $\texttt{p}_{\texttt{L}}$ is active, it will attempt to trigger the \failedinstance\ to excommunicate faulty processes. Nonetheless, it cannot collect sufficient acknowledgement after step (3) of the \failedinstance\ protocol, and hence shall identify itself as being isolated, thereby haling by committing suicide. This further increases $f$, and $\texttt{p}_{\texttt{L}}$ now has become faulty. Subsequently, the remaining correct processes will trigger the leader election, and eventually halt for not being able to collect a sufficient quorum of responses to their $\texttt{RequestVote}$, shutting down the application.

\vspace{2mm}
\noindent{\textbf{Case}} $({i > \lfloor \frac{n-1}{2}} \rfloor )$:
If $\texttt{p}_{\texttt{L}}$ identifies itself as being isolated (i.e., $\texttt{p}_{\texttt{L}}$ is among the $i$ isolated processes), it will halt and thereby becomes faulty. Once  $\texttt{p}_{\texttt{L}}$ is faulty, the remaining isolated processes will trigger the leader election, and will eventually halt for not being able to collect a sufficient quorum of responses to their $\texttt{RequestVote}$. In another words, the $i$ isolated processes will eventually become faulty, rendering $f > \lfloor \frac{n-1}{2} \rfloor $ and causing the application to shut down, as discussed earlier.

If $\texttt{p}_{\texttt{L}}$ is active, and can successfully drives the followers to complete all \newinstance\ and \failedinstance\ protocols that it triggers, the application remains operational. We note that whenever there is a new process added to the system membership list, or an existing process is excluded, the value of $n$ changes, which may moves the application out of or into anarchy.

\vspace{2mm}
\noindent{\textbf{Case}} $({f < \lfloor \frac{n-1}{2} \rfloor \wedge i < \lfloor  \frac{n-1}{2} \rfloor  \wedge (f + i)  > \lfloor \frac{n-1}{2}} \rfloor )$: 
If $\texttt{p}_{\texttt{L}}$ is isolated, it will halt and thereby become faulty. Once  $\texttt{p}_{\texttt{L}}$ is faulty, the remaining isolated processes will trigger the leader election.
Nonetheless, being isolated, they cannot collect a sufficient quorum of responses to their $\texttt{RequestVote}$. Consequently, they will halt, and thus become faulty. 
At this point,  the total number of faulty processes exceeds $\lfloor \frac{n-1}{2} \rfloor $, causing the application to shut down.

If $\texttt{p}_{\texttt{L}}$ is active, and can successfully drives the correct followers to exclude the faulty processes from their \texttt{peerLists} via the \failedinstance\ protocol, the application continues to be operational. Similar to the previous case where $i > \lfloor \frac{n-1}{2} \rfloor$, each process commission or excommunication changes the value of $n$, which may move the application out of or into anarchy.

\subsection{Timeout Configurations}
\label{subsec:timeout_config}

It should be clear from our security arguments thus far that safety of \codename\ does not depend on timing. In another words, \texttt{peerLists} of all active process converge even if they do not share any global clock, and their execution speed may differ. Nonetheless, its liveness (i.e., the service being able to make progress) is inevitably dependent upon timing, especially \heartbeatexchange\ and Leader Election protocols. A necessary condition for liveness in \codename\ is:

\begin{center}
\small
\texttt{broadcastTime} $\ll$ heartbeat timeout $\leq$ candidature timeout $\ll$ \texttt{EIATime} \\

\end{center} 

\noindent where \texttt{broadcastTime} is the the average time required for a process to send messages (in parallel) to other processes and receive their responses; heartbeat timeout and candidature timeout are the timeouts described in Section~\ref{subsec:overview}; and \texttt{EIATime} are events' inter-arrival time, with events being a new process instantiating, or failure/dismantling of existing processes in the membership cluster. Ideally, the timeouts should be an order of magnitude larger than the \texttt{broadcastTime}, and \texttt{EIATime} should be a few orders of magnitude larger than the timeouts. It is also reasonable to configure the heartbeat timeout to be less than or equal to  the candidature timeout. This setting allows the candidate to retry a leader election with a new \texttt{RequestVote} at a higher term, in case there is a split vote and the current term ends with no leader being elected.

While  \texttt{broadcastTime} and  \texttt{EIATime} depends on the underlying communication system and dynamic workloads of the application, respectively, the timeouts can be tuned at our discretion. As we show in our empirical evaluation, \texttt{broadcastTime} ranges from $10ms$ to $60ms$ for processes located across different geographical regions. We expect typical \texttt{EIATime} in typical applications to be in the neighborhood of tens of seconds. Consequently, the timeouts could be set in the range of $250-500ms$. Section~\ref{sec:eval} elaborates on the effect of timeout configuration on the performance of \codename.

\section{Implementation}
\label{sec:implementation}

To facilitate the incorporation of \codename's logic in the implementation of an enclave application \texttt{Prog}, we provide application developers with a \codename\ library. The library envelops the membership service's operations in a {\it control thread}, separating them from the original operational logic of \texttt{Prog} that is to be implemented in a {\it worker thread} (or threads). As a result, each \texttt{Prog}'s enclave process comprises one control thread, and one or more worker threads (as depicted in Figure~\ref{fig:implementation}). Our \codename\ library implementation contains approximately $3,000$ lines of C++ code, and builds on the Raft implementation by RethinkDB~\cite{rethinkdb}.

\definecolor{blizzardblue}{rgb}{0.67, 0.9, 0.93}
\definecolor{blanchedalmond}{rgb}{1.0, 0.92, 0.8}
\begin{figure}[t]
\centering  
\begin{tikzpicture}[every text node part/.style={align=center}]
\node[rectangle, draw = black, thick, dashed, fill = blizzardblue](t11){Control Thread};
\node[rectangle, draw = black, thick, dashed, fill = blanchedalmond, below of = t11, node distance = 0.6cm
](t12){Worker Thread};
\draw [draw=black] ([yshift=0.2cm,xshift=-0.2cm]t11.north west) rectangle ([yshift=-0.2cm,xshift=0.2cm]t12.south east);
\node[circle, draw = none, above of = t11, node distance = 0.6cm, xshift = -1cm] {$\texttt{p}_1$};

\node[rectangle, draw = black, thick, dashed, fill = blizzardblue, left of = t11, node distance = 2cm, yshift = -2.5cm](t21){Control Thread};
\node[rectangle, draw = black, thick, dashed, fill = blanchedalmond, below of = t21, node distance = 0.6cm
](t22){Worker Thread};
\draw [draw=black] ([yshift=0.2cm,xshift=-0.2cm]t21.north west) rectangle ([yshift=-0.2cm,xshift=0.2cm]t22.south east);
\node[circle, draw = none, above of = t21, node distance = 0.6cm, xshift = -1cm] {$\texttt{p}_2$};

\node[rectangle, draw = black, thick, dashed, fill = blizzardblue, right of = t11,node distance = 2cm, yshift = -2.5cm](t31){Control Thread};
\node[rectangle, draw = black, thick, dashed, fill = blanchedalmond, below of = t31, node distance = 0.6cm
](t32){Worker Thread};
\draw [draw=black] ([yshift=0.2cm,xshift=-0.2cm]t31.north west) rectangle ([yshift=-0.2cm,xshift=0.2cm]t32.south east);
\node[circle, draw = none, above of = t31, node distance = 0.6cm, xshift = -1cm] {$\texttt{p}_3$};

\draw[latex-latex, thick](t11.west) -- (t21.north);
\draw[latex-latex, thick](t11.east) -- node[yshift = -0.2cm, xshift = 1.15cm]{secure channel \\ via attestation} (t31.north);
\draw[latex-latex, thick](t31.west) -- (t21.east);

\end{tikzpicture}
\caption{A process in \codename\ comprises a control thread that is responsible for partaking in the autonomous membership service, and a worker thread that is responsible for handling the application's operational logic.}
\vspace{-2mm}
\label{fig:implementation}
\end{figure}
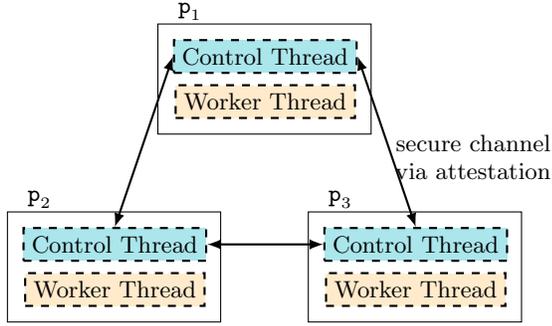

\vspace{2mm}
\noindent\textbf{Control thread vs worker thread.} 
The control thread is responsible for exchanging \texttt{heartbeat} messages, conducting remote attestation and secure channel establishment in the \newinstance\ protocol, and handling messages that are exchanged in other protocols of \codename. 
It is worth noting that the message exchange of the control thread entails context switching of the enclave. To reduce context switching overhead, our implementation piggybacks control thread's messages on the worker thread's interface with the untrusted application (e.g., its ECall and OCall) whenever necessary. 

Recall that both the control thread and the worker thread belong to the same enclave, the control thread can seal the application data handled by the worker thread to persistent storage, if need be, in the \shutdown\ protocol. In addition, the control thread handles the process's suicide by destroying its own enclave.

\vspace{2mm}
\noindent\textbf{Reducing IAS access overhead.} In the \newinstance\ protocol, the newly instantiated process has to attest itself to the leader, so as to convince the latter of its correct instantiation. If the new process is not located in the same physical host with the leader, they have to engage in a remote attestation procedure. In such a situation, the leader (i.e., remote verifier) cannot verify the attestation of an attesting process locally, but has to rely on the Intel's Attestation Service (IAS) to verify the signature contained in the attestation~\cite{ias}. Contacting the IAS in every run of \newinstance\ protocol would incur a significant overhead that is unfavorable for the application performance.

We make an observation that the multitenancy nature of cloud computing allows for a remedy of the aforementioned IAS access overhead. In particular, it is often the case that the CSP provisions many processes (either of the same or different applications) on the same physical host. Therefore, we can reduce the IAS access overhead by introducing a \texttt{HostAttestationDelegate} (\texttt{HAD}) enclave at each physical host, and routing the remote attestation between two processes instantiated on two separate physical hosts via their prospective \texttt{HAD} enclaves.

Let us consider two physical hosts $\texttt{h}_1$ and $\texttt{h}_2$, and denote their \texttt{HAD} enclaves by $\texttt{HAD\_E}_1$ and $\texttt{HAD\_E}_2$, respectively, as depicted in Figure~\ref{fig:HAD}. These two \texttt{HAD} enclaves engage in a conventional remote attestation, and rely on IAS to verify the attestation. Once $\texttt{HAD\_E}_1$ and $\texttt{HAD\_E}_2$ have completed the attestation and established a secure channel, any pair of processes $\texttt{p}^{\texttt{A}}_1$ and $\texttt{p}^{\texttt{A}}_2$ of an application \texttt{A} provisioned on  $\texttt{h}_1$ and $\texttt{h}_2$, respectively, can remotely attest one another without having to invoke the IAS. In particular, to attest its correct instantiating to $\texttt{p}^{\texttt{A}}_2$,
$\texttt{p}^{\texttt{A}}_1$ first leverages the local attestation procedure to prove its correctness to $\texttt{HAD\_E}_1$ which then conveys this attestation result to $\texttt{HAD\_E}_2$ via the secure channel. Finally, $\texttt{HAD\_E}_2$ locally attests to $\texttt{p}^{\texttt{A}}_2$ that it is correctly instantiated, gaining the latter's trust, and then conveys $\texttt{p}^{\texttt{A}}_1$'s attestation result to $\texttt{p}^{\texttt{A}}_2$. This approach essentially amortizes the IAS access overhead incurred by the attestation of \texttt{HAD} enclaves, and allows two processes  $\texttt{p}^{\texttt{A}}_1$ and $\texttt{p}^{\texttt{A}}_2$ of an application \texttt{A} provisioned on two separate hosts $\texttt{h}_1$ and $\texttt{h}_2$ to conduct remote attestation without contacting the IAS. 

\definecolor{aquamarine}{rgb}{0.5, 1.0, 0.83}
\definecolor{cream}{rgb}{1.0, 0.99, 0.82}
\begin{figure}[t]
\centering  
\begin{tikzpicture}[every text node part/.style={align=center}]
\node[rectangle, draw = black, fill = cream](hade1){$\texttt{HAD\_E}_1$};
\node[rectangle, draw = black, fill = aquamarine, right of = hade1, node distance = 1.5cm
](pA1){$\texttt{p}^{\texttt{A}}_{1}$};
\node[rectangle, draw = black, fill = aquamarine, right of = pA1, node distance = 0.7cm
](pB1){$\texttt{p}^{\texttt{B}}_{1}$};
\node[rectangle, draw = black, fill = aquamarine, right of = pB1, node distance = 0.7cm
](pC1){$\texttt{p}^{\texttt{C}}_{1}$};
\node[rectangle, draw = black, fill = aquamarine, right of = pC1, node distance = 0.7cm
](pD1){$\texttt{p}^{\texttt{D}}_{1}$};

\draw [draw=blue] ([yshift=0.2cm,xshift=-0.2cm]hade1.north west) rectangle ([yshift=-0.15cm,xshift=0.2cm]pD1.south east);

\node[rectangle, draw=none, above of = hade1, xshift = -0.5cm, node distance = 0.6cm](host1){$\texttt{h}_1$};

\node[rectangle, draw = black, fill = cream, below of = hade1, node distance = 1.6cm](hade2){$\texttt{HAD\_E}_2$};
\node[rectangle, draw = black, fill = aquamarine, right of = hade2, node distance = 1.5cm
](pA2){$\texttt{p}^{\texttt{A}}_{2}$};
\node[rectangle, draw = black, fill = aquamarine, right of = pA2, node distance = 0.7cm
](pB2){$\texttt{p}^{\texttt{B}}_{2}$};
\node[rectangle, draw = black, fill = aquamarine, right of = pB2, node distance = 0.7cm
](pC2){$\texttt{p}^{\texttt{C}}_{2}$};
\node[rectangle, draw = black, fill = aquamarine, right of = pC2, node distance = 0.7cm
](pD2){$\texttt{p}^{\texttt{D}}_{2}$};
\draw [draw=blue] ([yshift=0.2cm,xshift=-0.2cm]hade2.north west) rectangle ([yshift=-0.15cm,xshift=0.2cm]pD2.south east);

\node[rectangle, draw=none, above of = hade2, xshift = -0.5cm, node distance = 0.6cm](host2){$\texttt{h}_2$}; 

\draw[latex-latex, thick](hade1.south) -- (hade2.north);

\draw[latex-latex, dashed, thick]  (hade1.east) -- (pA1.west);
\draw[latex-latex, dashed, thick]  ($(pB1.north)+(0,0)$) -- ($(pB1.north)+(0,+0.3)$) -| ($(hade1.north)+(0.2,0)$){};
\draw[latex-latex, dashed, thick]  ($(pC1.north)+(0,0)$) -- ($(pC1.north)+(0,+0.5)$) -| (hade1.north){};
\draw[latex-latex, dashed, thick]  ($(pD1.north)+(0,0)$) -- ($(pD1.north)+(0,+0.7)$) -| ($(hade1.north)+(-0.2,0)$){};

\draw[latex-latex, dashed, thick]  (hade2.east) -- (pA2.west);
\draw[latex-latex, dashed, thick]  ($(pB2.south)+(0,0)$) -- ($(pB2.south)+(0,-0.3)$) -| ($(hade2.south)+(0.2,0)$){};
\draw[latex-latex, dashed, thick]  ($(pC2.south)+(0,0)$) -- ($(pC2.south)+(0,-0.5)$) -| (hade2.south){};
\draw[latex-latex, dashed, thick]  ($(pD2.south)+(0,0)$) -- ($(pD2.south)+(0,-0.7)$) -| ($(hade2.south)+(-0.2,0)$){};

\draw[latex-latex, dashed, thick]  ($(pD1.east)+(0.5,-0.25)$) -- node[]{Local \\ Attestation} ($(pD1.east)+(2.5,-0.25)$);
\draw[latex-latex, thick]  ($(pD2.east)+(0.5,0.25)$) -- node[]{Remote \\ Attestation} ($(pD2.east)+(2.5,0.25)$);

\end{tikzpicture}
\caption{\codename\ introduces \texttt{HostAttestationDelegate} enclave at each host, and leverages  them to reduce IAS access overhead incurred in the \newinstance\ protocol.}
\vspace{-2mm}
\label{fig:HAD}
\end{figure}
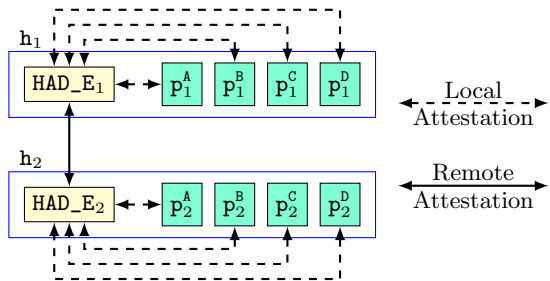

\section{Evaluation}
\label{sec:eval}

In this section, we report our experimental study of \codename. Our experiments are conducted in two different settings.  
The first setting is based on an in-house (local) cluster consisting of 35 servers, each equipped with Intel Xeon E3-1240 2.1GHz CPUs, 32GB RAM and 2TB hard drive. The latency between any two servers are roughly $0.13ms$. In this setting, we run each process on a separate server. The second setting is based on  Google Cloud Platform (GCP), in which we use a separate instance provisioned with  2 vCPUs and 8GB RAM for each process. The instances on GCP span across 4 regions, namely Oregon (us-west1), Los Angeles (us-west2), South Carolina (us-east1) and North Virginia (us-east4). The average latency between these regions observed in our experiment is detailed in Table~\ref{tab:GCP_latency}.

The trusted code base in our experiment is implemented using Intel SGX SDK~\cite{sgx_sdk}. 
For the experiments that we run on GCP, we configured the SDK to run in simulation mode, as SGX is not available on GCP instances. We measured the latency of each SGX operation on our local cluster's CPU with SGX Enabled BIOS support, and injected it to the simulation. We observe that public key operations are expensive: signing and signature verification take roughly $450 \mu s$ and $844 \mu s$, respectively. Context switching and symmetric key operations take less than $5 \mu s$. Remote attestation, which involves access to the IAS, takes about $250 ms$ on average. Unless otherwise stated, the results reported in the following are averaged over $20$ independent runs.

\definecolor{grad0}{RGB}{235, 130, 128} 
\definecolor{grad1}{RGB}{240, 185, 147} 
\definecolor{grad2}{RGB}{245, 226, 150} 
\definecolor{grad3}{RGB}{225, 235, 185} 
\definecolor{grad4}{RGB}{220, 245, 215} 

\begin{table} \centering
\caption{Latency (ms) between different regions on GCP.}
\label{tab:GCP_latency}
\resizebox{0.45\textwidth}{!} {
\begin{tabular}{|l|r|r|r|r|}
\hline
\textbf{Zone} & \textbf{us-west1} & \textbf{us-west2} & \textbf{us-east1} & \textbf{us-east4}\\
\hline\hline
\textbf{us-west1} & 0.0 & \cellcolor{grad3} 24.7 & \cellcolor{grad1} 66.7 & \cellcolor{grad2} 59.0 \\
\hline
\textbf{us-west2} & \cellcolor{grad3} 24.7 & 0.0 & \cellcolor{grad1} 62.9 & \cellcolor{grad1} 60.5 \\
\hline
\textbf{us-east1} & \cellcolor{grad1} 66.7 & \cellcolor{grad1} 62.9 & 0.0 & \cellcolor{grad4} 12.7 \\
\hline
\textbf{us-east4} & \cellcolor{grad2} 59.1 & \cellcolor{grad1} 60.4 & \cellcolor{grad4} 12.7 & 0.0 \\
\hline

\end{tabular}
}
\end{table}

\subsection{\codename\ Overhead}
%

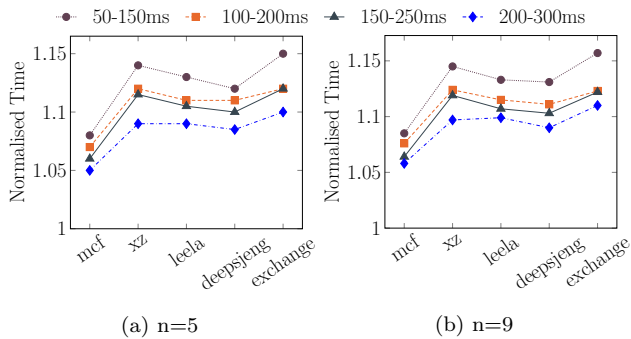
\begin{figure}[t]
\centering
\begin{subfigure}{.23\textwidth}
\begin{tikzpicture}[thick, scale = 0.45]
\begin{axis}[
	ticklabel style = {font=\LARGE},
    ylabel={Normalised Time},  
    x label style={font = \LARGE},
    y label style={font = \LARGE, at={(-0.05,0.5)}},
    xticklabels={mcf, xz, leela, deepsjeng,   exchange},
    xtick=data,
    ymin = 1.0,
    xticklabel style={rotate=35},
    legend style={at={(-0.08,1.1)},anchor=west, draw=none,legend columns=4, font = \LARGE, column sep = 0.4cm},
]

\addplot [color = eggplant, mark=*,mark options={scale=1.5, solid, fill=eggplant}, densely dotted, thick] coordinates {
(1,1.08) 
(2,1.14) 
(3,1.13) 
(4,1.12)
(5,1.15)
};

\addplot [color = deepcarrotorange, mark=square*,  mark options={scale=1.5,solid,fill=deepcarrotorange}, densely dashed] coordinates { 
(1,1.07) 
(2,1.12) 
(3,1.11) 
(4,1.11)
(5,1.12)
};

\addplot [color = charcoal, mark=triangle*,  mark options={scale=2,solid,fill=charcoal}] coordinates { 
(1,1.06) 
(2,1.115) 
(3,1.105) 
(4,1.10)
(5,1.12)
};

\addplot [color = blue, mark=diamond*,  mark options={scale=2,solid,fill=blue}, dashdotted] coordinates { 
(1,1.05) 
(2,1.09) 
(3,1.09) 
(4,1.085)
(5,1.10)
};

\legend{50-150ms, 100-200ms, 150-250ms, 200-300ms}
\end{axis}
\end{tikzpicture}
\caption{n=5}
\label{subfig:overhead_5}
\end{subfigure}
\begin{subfigure}{.23\textwidth}
\vspace{4mm}
\begin{tikzpicture}[thick, scale = 0.45]
\begin{axis}[
	ticklabel style = {font=\LARGE},
    ylabel={Normalised Time},  
    x label style={font = \LARGE},
    y label style={font = \LARGE, at={(-0.05,0.5)}},
    xticklabels={mcf, xz, leela, deepsjeng,   exchange},
    xtick=data,
    ymin = 1.0,
    xticklabel style={rotate=35},
    legend style={at={(-0.08,1.1)},anchor=west, draw=none,legend columns=4, font = \LARGE, column sep = 0.4cm},
]

\addplot [color = eggplant, mark=*,mark options={scale=1.5, solid, fill=eggplant}, densely dotted, thick] coordinates {
(1,1.085) 
(2,1.145) 
(3,1.133) 
(4,1.131)
(5,1.157)
};

\addplot [color = deepcarrotorange, mark=square*,  mark options={scale=1.5,solid,fill=deepcarrotorange}, densely dashed] coordinates { 
(1,1.076) 
(2,1.124) 
(3,1.115) 
(4,1.111)
(5,1.123)
};

\addplot [color = charcoal, mark=triangle*,  mark options={scale=2,solid,fill=charcoal}] coordinates { 
(1,1.064) 
(2,1.119) 
(3,1.107) 
(4,1.103)
(5,1.122)
};

\addplot [color = blue, mark=diamond*,  mark options={scale=2,solid,fill=blue}, dashdotted] coordinates { 
(1,1.058) 
(2,1.097) 
(3,1.099) 
(4,1.09)
(5,1.11)
};

\end{axis}
\end{tikzpicture}
\caption{n=9}
\label{subfig:overhead_9}
\end{subfigure}
\caption{\codename's overhead compared to vanilla SGX execution. The running time of each benchmark is normalized against its own vanilla SGX execution's.}
\label{fig:cpu_spec_overhead}
\end{figure}

We evaluate overhead incurred by \codename\ over vanilla SGX execution using five benchmarks (i.e., \textit{mcf, deepsjeng, leela, exchang2}, and \textit{xz}) selected from SPEC CPU2017~\cite{spec_cpu} on our local cluster.
We leverage Intel SGX SDK~\cite{sgx_sdk} to port the benchmarks into enclave execution, enveloping their operational logic in a worker thread.  We then inject a control thread that is responsible to implement the membership service to the enclave using our \codename\ library. 

Figure~\ref{fig:cpu_spec_overhead} reports the running time of each \codename-enabled benchmarks under different range of heartbeat timeouts  normalized against their own vanilla SGX execution's running time.
For a given range $[a,b]$, a heartbeat timeout is chosen uniformly at random between $a$ and $b$, inclusively. The overhead ranges from $5\%$ to $16\%$, with smaller timeout range leads to higher overhead. This is so because smaller heartbeat timeouts result in higher frequency of the leader contacting the follower processes to maintain its authority, which in turns leads to higher communication overhead and control switching. We also observe that the overhead is more evident when benchmark workload is computation intensive and requires  few control switching for I/O (e.g., for the heartbeat timeout range of $[50-100](ms)$, {\it mcf} witnesses an overhead of $8\%$, whereas {\it exchange} sports a $15\%$ overhead. 
This makes sense because our implementation piggybacks \codename\ control thread's messages on the worker thread's interface with the untrusted application (e.g., its ECall and OCall) whenever necessary. 
Thus, the more interactions the worker thread has with the untrusted application over the course of the execution, the fewer control switching the process has to trigger exclusively for exchanging the control thread's messages, and hence lower the overhead. Comparing across figure~\ref{subfig:overhead_5} and figure~\ref{subfig:overhead_9}, we can see that while the overhead increases when there are more processes in the system, the difference is insignificant. In particular, when $n$ increases from $5$ to $9$, the difference in the normalized running time is less than $1\%$ for all benchmarks in our experiment.

\subsection{Leader Election}
In the second set of experiments, we study the effect of timeout configurations on \codename\ leader election. We measure the leader election latency as a period from when we crash the leader of a cluster of $n$ servers (with $n$ ranges from $5$ to $33$) to the moment when the new leader announces its authority. We experiment with different range of heartbeat timeouts and candidature timeout. For a given range $[a,b]$, a heartbeat timeout is chosen uniformly at random between $a$ and $b$, inclusively, while the candidature timeout is set to $5 \times b$.

\begin{figure}[t]
\centering
\begin{subfigure}{.23\textwidth}
\begin{tikzpicture}[thick, scale = 0.45]
\begin{axis}[
	ticklabel style = {font=\LARGE},
    xtick=data,
    xticklabels={$5$,$9$,$17$, $33$},
    xlabel={$n$},
    ymax = 300,
    ylabel={Latency (ms)},
    x label style={font = \LARGE},
    y label style={font = \LARGE, at={(-0.05,0.5)}},
    legend style={at={(-0.08,1.1)},anchor=west, draw=none,legend columns=4, font = \LARGE, column sep = 0.4cm},
]

\addplot [color = eggplant, mark=*,mark options={scale=1.5, solid, fill=eggplant}, densely dotted, thick] coordinates {
(1,68) 
(2,71) 
(3,76) 
(4,78) 
};

\addplot [color = deepcarrotorange, mark=square*,  mark options={scale=1.5,solid,fill=deepcarrotorange}, densely dashed] coordinates { 
(1,145) 
(2,152) 
(3,161) 
(4,163) 

};

\addplot [color = charcoal, mark=triangle*,  mark options={scale=2,solid,fill=charcoal}] coordinates { 
(1,189) 
(2,190) 
(3,195) 
(4,201) 
};

\addplot [color = blue, mark=diamond*,  mark options={scale=2,solid,fill=blue}, dashdotted] coordinates { 
(1,238) 
(2,246) 
(3,251) 
(4,253) 
};

\legend{50-150ms, 100-200ms, 150-250ms, 200-300ms}
\end{axis}
\end{tikzpicture}
\caption{Local Cluster}
\end{subfigure}
\begin{subfigure}{.23\textwidth}
\vspace{4.2mm}
\begin{tikzpicture}[thick, scale = 0.45]
\begin{axis}[
	ticklabel style = {font=\LARGE},
    xtick=data,
    xticklabels={$5$,$9$,$17$, $33$},
    ymin = 50,
    xlabel={$n$},
    ylabel={Latency (ms)},
    x label style={font = \LARGE},
    y label style={font = \LARGE, at={(-0.05,0.5)}},
   legend style={at={(0.18,1.1)},anchor=west, draw=none,legend columns=4, font = \LARGE, column sep = 0.2cm},
]
\addplot [color = eggplant, mark=*,mark options={scale=1.5, solid, fill=eggplant}, densely dotted, thick] coordinates {
(1,405) 
(2,456) 
(3,489) 
(4,552) 
};

\addplot [color = deepcarrotorange, mark=square*,  mark options={scale=1.5,solid,fill=deepcarrotorange}, densely dashed] coordinates { 
(1,375) 
(2,389) 
(3,438) 
(4,497)
};

\addplot [color = charcoal, mark=triangle*,  mark options={scale=2,solid,fill=charcoal}] coordinates { 
(1,359) 
(2,368) 
(3,402) 
(4,452) 
};

\addplot [color = blue, mark=diamond*,  mark options={scale=2,solid,fill=charcoal}, dashdotted] coordinates { 
(1,325) 
(2,386) 
(3,426) 
(4,458) 
};

\end{axis}
\end{tikzpicture}
\caption{GCP}
\label{subfig:effect_of_q1_q2}
\end{subfigure}
\vspace{-2mm}
\caption{Leader election latency.}
\vspace{-2mm}
\label{fig:leader_election_latency}
\end{figure}
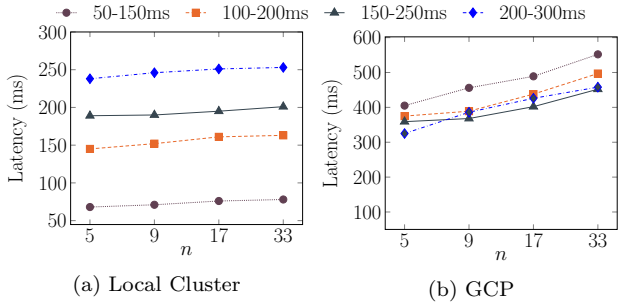

Figure~\ref{fig:leader_election_latency} reports the latency under difference timeout settings on our local cluster and GCP. On our local cluster, the smaller the timeout range, the lower the leader election latency is. However, this phenomenon is not observed in our experiments on the GCP. More specifically, with the timeout range of $[50-100](ms)$, the leader election latency on GCP ranges from $405-552ms$, while the timeout range of $[200-300](ms)$ yields the lower leader election latency ranging from $325-458ms$. 

A careful analysis of the difference between our local cluster and GCP confirms the necessity of proper timeout configurations (see Section~\ref{subsec:timeout_config}) with respect to communication latency (or \texttt{broadcastTime}). Recall that on our local cluster, the latency between any two servers is roughly $0.13ms$, whereas that of GCP ranges from $24ms$ to $66ms$, depends on the geolocation of the servers. When the heartbeat and candidature timeout is too small compared to the communication latency, the candidates face difficulty in collecting sufficient responses to its $\texttt{RequestVote}$ in time, entering the new terms and triggering new leader election prematurely. This leads to unnecessary leader elections, and exacerbates the system availability.


\subsection{\newinstance\ Protocol}

\definecolor{cornellred}{rgb}{0.7, 0.11, 0.11}
\definecolor{arsenic}{rgb}{0.23, 0.27, 0.29}
\definecolor{auburn}{rgb}{0.43, 0.21, 0.1}
\definecolor{charcoal}{rgb}{0.21, 0.27, 0.31}
\definecolor{deepcarrotorange}{rgb}{0.91, 0.41, 0.17}
\definecolor{eggplant}{rgb}{0.38, 0.25, 0.32}

\begin{figure}[t]
\centering
\begin{subfigure}{.23\textwidth}
\begin{tikzpicture}[thick, scale = 0.45]
\begin{axis}[
    width = 2\textwidth,
	height= \axisdefaultheight,
	ticklabel style = {font=\LARGE},
    ybar,
    enlarge x limits=0.2,
    legend style={at={(.35,1.1)},anchor=west, draw=none,legend columns=2, font = \LARGE, column sep=0.2cm},
    bar width=15pt,  
    area legend,
    ylabel={Latency (ms)},   
    xlabel={$n$},   
    y label style={font = \LARGE, at={(-0.05,0.5)}},   
    x label style={font = \LARGE},   
    xticklabels={5, 9, 17, 33},
    xtick=data,
    ymin = 0,
    ymax = 550
]

\addplot[color = eggplant,thick, pattern color = eggplant,pattern=north west lines] coordinates {
(1,270) 
(2,275) 
(3,290) 
(4,315) 
};

\addplot [color = deepcarrotorange, pattern color = deepcarrotorange, pattern = crosshatch dots] coordinates { 
(1,18) 
(2,25) 
(3,45) 
(4,65) 

};

\legend{w/o \texttt{HAD} enclaves, w \texttt{HAD} enclaves}
\end{axis}
\end{tikzpicture}
\caption{Local Cluster}
\label{subfig:new_process_LC}
\end{subfigure}
\begin{subfigure}{.23\textwidth}
\vspace{4.2mm}
\begin{tikzpicture}[thick, scale = 0.45]
\begin{axis}[
    width = 2\textwidth,
	height= \axisdefaultheight,
	ticklabel style = {font=\LARGE},
    ybar,
    enlarge x limits=0.2,
    legend style={at={(.2,1.1)},anchor=west, draw=none,legend columns=2, font = \LARGE, column sep=0.2cm},
    bar width=15pt,  
    area legend,
    ylabel={Latency (ms)},   
    xlabel={$n$},   
    y label style={font = \LARGE, at={(-0.05,0.5)}},   
    x label style={font = \LARGE},   
    xticklabels={5, 9, 17, 33},
    xtick=data,
    ymin = 0,
    ymax = 550
]

\addplot[color = eggplant,thick, pattern color = eggplant,pattern=north west lines] coordinates {
(1,380) 
(2,450) 
(3,495) 
(4,538) 
};

\addplot [color = deepcarrotorange, pattern color = deepcarrotorange, pattern = crosshatch dots] coordinates { 
(1,145) 
(2,210) 
(3,260) 
(4,285) 

};

\end{axis}
\end{tikzpicture}
\caption{GCP}
\label{subfig:new_process_GCP}
\end{subfigure}
\vspace{-2mm}
\caption{\newinstance's latency.}
\vspace{-4mm}
\label{fig:new_process_lt}
\end{figure}
We now examine the cost of the \newinstance\ protocol with and without \texttt{HostAttestationDelegate} (\texttt{HAD}) enclave. Recall that by incorporating \texttt{HAD} enclave, the leader and the new process can engage in a remote attestation without contacting the IAS. 

Figure~\ref{fig:new_process_lt} depicts the latency of \newinstance\ protocol with respect to different number of existing processes in the system at the beginning of the new process's commission. The latency of \newinstance\ protocol is measured as the duration between the new process $\texttt{p}_{\texttt{j}}$ signaling its join request till $\texttt{p}_{\texttt{j}}$ receiving the application secret and becoming fully functional. 

On our local cluster (Figure~\ref{subfig:new_process_LC}), \newinstance\ without \texttt{HAD} enclaves takes from $270-325ms$, whereas incorporating \texttt{HAD} enclaves reduces such a cost to $18-65$ms. This difference is primarily due to the cost of IAS access, which is approximately $250$ms in our experiments. A similar result is observed on GCP experiments (Figure~\ref{subfig:new_process_GCP}), wherein incorporating \texttt{HAD} enclaves significantly reduces the cost of the \newinstance\ protocol. Compared to the experiments on our local cluster, the \newinstance\ latency observed on GCP is higher (e.g., $145-
285ms$ compared to $18-65ms$ with \texttt{HAD} enclaves). This is so because the cost of the underlying consensus protocol itself, especially the broadcast time during the consensus run, differ significantly between the two settings.

\subsection{\failedinstance\ Protocol}
\definecolor{armygreen}{rgb}{0.29, 0.33, 0.13}
\definecolor{auburn}{rgb}{0.8, 0.5, 0.2}
\begin{figure}[t]
\centering
\begin{tikzpicture}[thick, scale = 0.45]
\begin{axis}[
	ticklabel style = {font=\LARGE},
    ybar,
    enlarge x limits=0.2,
    legend style={at={(.0,1.1)},anchor=west, draw=none,legend columns=2, font = \LARGE, column sep=0.2cm},
    bar width=18pt,  
    area legend,
    ylabel={Latency (ms)},   
    xlabel={$n$},   
    y label style={font = \LARGE, at={(-0.05,0.5)}},   
    x label style={font = \LARGE},   
    xticklabels={5, 9, 17, 33},
    xtick = data,
    ymin = 0,
    ]

\addplot[color = armygreen,thick, pattern color = armygreen,pattern=north east lines] coordinates {
(1,52) 
(2,60) 
(3,73) 
(4,91) 
};

\addplot [color = auburn, pattern color = auburn, pattern = grid] coordinates { 
(1,350) 
(2,428) 
(3,505) 
(4,553) 

};

\legend{Local Cluster, GCP}
\end{axis}
\end{tikzpicture}
\caption{\failedinstance's latency.}
\vspace{-4mm}
\label{fig:remove_process_lt}
\end{figure}

Finally, we examine the cost of the \failedinstance\ protocol by crashing a process $\texttt{p}_{\texttt{h}}$, and measure the latency from such the crash till the time when $\texttt{p}_{\texttt{h}}$ is expelled from the leader's $\texttt{peerList}$. 
On our local cluster, the heartbeat timeout is chosen randomly in the range of $[25-50]$ms, whereas the corresponding range on GCP is $[200-300]ms$. Figure~\ref{fig:remove_process_lt} plots the latency of the \failedinstance\ protocol observed on our local cluster and GCP, with respect to different $n$. 

Similar to \newinstance\ protocol, the latency observed on GCP is significantly higher than that observed in our local cluster. We attribute this to the difference in the timeout configurations, which in turn is influenced by the broadcast time of each setting.

\section{Related work}
\label{sec:related_work}
\noindent\textbf{Membership service.}
Birman et al.~\cite{birman_membership} formulated the problem of group membership in asynchronous environments. The authors acknowledge that the system's liveness entails the possibility of erroneous failure detection, and the need of suppressing communication from a process that is unjustly excommunicated. Our autonomous membership service follows this observation, fulfilling the excommunication by having an unjustly excommunicated process committing suicide. Amir et al.~\cite{transis} proposed a membership protocol that relies on atomic broadcast, whereas Mishra et al. ~\cite{membership_partial_order} leveraged a multicast facility that preserves the partial order of messages exchanged among the communicating processes to construct a membership protocol. Our approach, on the other hand, does not make any assumption on the underlying broadcast/multicast primitives. Instead, it builds on a well-known consensus protocol, namely Raft~\cite{raft}, to implement the membership service. 

\vspace{2mm}

\noindent\textbf{Live migration and replications of SGX enclaves.}
Gu et al.~\cite{sgx_migration} present technique for live enclave migration on the cloud. In contrary to our membership service, their system model allows only one instance of the application to be functional at any time. ReplicaTEE~\cite{replicatee} attempts to enable seamless replication of SGX enclaves by relying on a distributed directory service layer that handles provisioning of enclave processes, and keeps track of the number of existing processes. Nevertheless, it is unclear how the directory service's replicas are protected against forking attack, wherein the adversary splits the directory service replicas into two cliques, each of which is fully operational but unaware of the other. This attack effectively allows the adversary to replicate more enclave processes than what would otherwise be allowed by the SLA.

%

\section{Conclusion}
\label{sec:conclusion}

We have described an autonomous membership service for enclave-based applications. 
While the application bootstrap necessitates the involvement of the application owner, 
all subsequent process commission and decommission are administered the existing and active processes of the application. The proposed membership service necessarily admits unjust excommunication of a non-faulty process from the membership group, which is then fulfilled by the unjustly excommunicated process committing suicide. The experimental study shows that \codename\ incurs $5\% - 16\%$ overhead compared to vanilla enclave execution. 

\bibliographystyle{ACM-Reference-Format}
\bibliography{ms}

\end{document}